\documentclass[final,3p,times,twocolumn]{elsarticle}
\usepackage{graphicx}
\usepackage{amssymb}

\def\Journal#1#2#3#4{{#1} {\bf #2} (#3) #4.}

\def\NIM{Nucl.\,Inst. and Meth.}

\def\NP{Nucl.\,Phys.}

\def\PL{Phys.\,Lett.}

\def\PRL{Phys.\,Rev.\,Lett.}

\def\PR{Phys.\,Rev.}

\def\Ho{$^{163}$Ho}
\def\mn{$m_\nu$}

\def\mus{$\mu$s}

\def\de{$\Delta E$}
\def\fwhm{$_{\mathrm{FWHM}}$}

\journal{Astroparticle Physics}

\begin{document}

\begin{frontmatter}

\title{The Electron Capture Decay of $^{163}$Ho to Measure the Electron 
Neutrino Mass with sub-eV Accuracy}

\author[miami]{Massimiliano Galeazzi}
\author[genova]{Flavio Gatti}
\author[roma]{Maurizio Lusignoli}
\author[milano]{Angelo Nucciotti}
\author[milano]{Stefano Ragazzi}
\author[lisboa]{Maria Ribeiro Gomes}

\address[miami]{University of Miami, 1320 Campo Sano Dr., Coral Gables, FL 33146, USA}
\address[genova]{INFN-Genova and Universit\`{a} degli Studi di Genova, 
Via Dodecanneso 33, 16146 Genova, Italy}
\address[roma]{INFN-Roma and Universit\`{a} di Roma La Sapienza, 
Piazzale Aldo Moro 5, 00185 Roma, Italy}
\address[milano]{INFN-Milano Bicocca and Universit\`{a} di Milano Bicocca, 
Piazza della Scienza 3, 20126 Milano, Italy}
\address[lisboa]{Centro de Fisica Nuclear da Universidade de Lisboa, Av. Prof. Gama Pinto, 2, 1649-003, Lisboa, Portugal}

\begin{abstract}

We have investigated the possibility of measuring the
electron neutrino mass with sub-eV sensitivity by studying 
the electron capture decay of $^{163}$Ho with cryogenic microcalorimeters. 
In this paper we will introduce an experiment's concept, discuss
the technical requirements, and identify a roadmap to 
reach a sensitivity of 0.1 eV/$c^2$ and beyond.
\end{abstract}


\end{frontmatter}

\section{Introduction}
\label{introduction}

Assessing the neutrino mass scale is one of the major challenges in 
today particle physics and astrophysics. Although neutrino oscillation 
experiments have clearly shown that there are at least three neutrinos 
with different masses, their absolute values remain unknown. 
Neutrino flavor oscillations are sensitive to the difference between the 
squares of neutrino mass eigenvalues and have been measured by solar, 
atmospheric, reactor, and accelerator experiments \cite{oscillations}. 
Combining such results, however, does not lead to an absolute value for 
the eigenmasses, and a dichotomy between two possible scenarios, 
dubbed "normal" and "inverted" hierarchies, exists. 
The scenario could be complicated further by the possible existence of 
additional "sterile" neutrino eigenvalues at different mass scales \cite{SterileNu}. 

The value of the neutrino mass has many implications, from cosmology 
to the Standard Model of particle physics. In cosmology the neutrino mass affects the 
formation of large-scale structure in the universe. In particular, 
 neutrinos tend to damp structure clustering, before they have cooled sufficiently 
to become non-relativistic, 
with an effect that is dependent on their mass \cite{nu-structureformation}. 
In the framework of $\Lambda$-CDM cosmology, the scale dependence of 
clustering observed in the Universe can, indeed, be used to set an 
upper limit on the neutrino mass sum in the range between about 0.3 and 
2\,eV \cite{nu-CMB},  
although this value is strongly model-dependent. In 
particle physics, 
a determination of the absolute scale of neutrino masses would 
help shedding light on the mechanisms of mass generation. 

For years, laboratory experiments based on the study of proper 
nuclear processes have been used to directly measure the neutrino 
masses. In particular, single beta decay has been, historically, 
the traditional and most direct method to investigate the electron 
(anti)neutrino mass \cite{betaendpoint}. Neutrinoless double beta decay has also been 
suggested as a powerful tool to measure the electron neutrino mass, 
although the decay itself, and thus its efficacy at measuring the 
neutrino mass, is dependent on the assumption that 
the neutrino is, in fact, a Majorana particle \cite{bb-Majorana}.

To date, the study of the beta decay of $^3$H using electrostatic 
spectrometers has been the most sensitive approach, yielding an 
upper limit on the electron anti-neutrino mass of 2.2\,eV/$c^2$ \cite{MainzTroitsk}. 
In the near future, the new experiment KATRIN will analyze the $^3$H 
beta decay end-point with a much more sensitive electrostatic 
spectrometer and an expected statistical sensitivity of about 
0.2\,eV/$c^2$ \cite{KATRIN}.

The calorimetric measurement of the beta decay of $^{187}$Re using 
cryogenic microcalorimeters has also been successfully used \cite{renioExp}, 
and a new experiment, MARE, is planned to have a sensitivity for 
the neutrino mass comparable to that of KATRIN \cite{MARE}. 
Although a calorimetric experiment is not affected by the many 
systematic uncertainties that plagued electrostatic spectrometers 
in the past, the very nature of the experiments requires that 
all decays are being measured, not only those with energy near to the end-point.
 To maximize the useful 
 experimental statistics, a calorimetric 
experiment thus requires an isotope with very low end-point energy, 
hence the choice of $^{187}$Re, with an end-point energy of about 
2.5\,keV \cite{renioExp}. 

The potential of MARE is quite promising and its first phase is 
well under way. However, its full implementation and, more important, 
its extension beyond the 0.2\,eV/$c^2$ limit currently planned, are 
strongly affected by the long decay time of $^{187}$Re. 
With a half-life comparable to the age of the Universe -- about $43\times10^9$\,years \cite{renioExp} --,
the mass of radioactive material necessary to reach the high 
statistics required by the experiment in a reasonable amount of time 
puts serious constraints on the experimental design and fabrication \cite{Nucciotti-Sens}.

It was suggested thirty years ago that electron capture (EC) decays with 
low $Q$--values could be used as an alternative to single beta decay
for the direct determination of the electron neutrino mass. In 1981 a group at Princeton University \cite{Bennett:1981} tried to obtain a limit on the electron neutrino mass measuring the X-rays emitted after EC of electrons in different levels. In the same year  De R$\acute{\rm u}$jula \cite{DeRujula:1981ti} proposed to 
measure the spectra of photons emitted by inner brehmsstrahlung in EC decays 
(IBEC) with the K shell capture forbidden, and an experimental activity followed \cite{Jonson:1982gr}.
 The spectrum of emitted Auger
 electrons was  also considered  later \cite{DeRujula:1982bq}. The most appealing suggestion, 
that we will follow in this Letter, considers a calorimetric experiment, where 
all the de-excitation energy is recorded \cite{DeRujula:1982qt}.

The EC decay of $^{163}$Ho to $^{163}$Dy is the decay with the lowest 
known $Q$--value and its half--life of a few thousand 
years is much less than the $^{187}$Re one. Detailed theoretical calculations 
and sensitivity to the neutrino mass in the $^{163}$Ho case are
reported in \cite{DeRujula:1982qt}.
In 1997 the $^{163}$Ho decay was measured using cryogenic microcalorimeters 
\cite{GattiHo:1997} and in 2008 its use was suggested included
in the framework of the MARE experiment \cite{MARE-Ho:2008}. 
More recently, a similar, parallel effort with magnetic 
microcalorimeters has also started \cite{Ranitzsch}.
Unfortunately, at this time the 
experimental measurements of the $Q$--value range from about 
2.2\,keV to 2.8\,keV, an uncertainty that translates to a factor 3 to 4 on the
neutrino mass sensitivity achievable by a Holmium experiment. 

In this paper we discuss the potential of an experiment to measure the
electron neutrino mass with sub-eV sensitivity by studying the electron 
capture decay of $^{163}$Ho with cryogenic microcalorimeters. 
In particular, we will introduce the experiment's concept, discuss
the technical requirements, and identify a roadmap to 
reach a sensitivity of 0.1\,eV/$c^2$ and beyond.

\section{Measuring the electron neutrino mass 
through the $^{163}$Ho electron capture decay}
\label{section2}

There are at least three proposed independent methods to assess the 
neutrino mass from the $^{163}$Ho EC decay:
\begin{enumerate}
\item{Absolute M capture rates or M/N capture rate ratios \cite{Bennett:1981};}
\item{Inner Bremsstrahlung (IB) end-point \cite{DeRujula:1981ti};}
\item{Total (calorimetric) absorption spectrum end-point \cite{DeRujula:1982qt}.}
\end{enumerate}

\subsection*{Absolute M capture rates or M/N capture rate ratios}

All the experimental research so far has focused on the atomic 
emissions - photons and electrons - following the EC to exploit the 
possibility of constraining simultaneously the transition $Q$ and the neutrino 
mass from the relative probabilities of M and N shell capture or absolute M 
capture rate \cite{Bennett:1981, expHo,HN}.

The EC decay rate can be expressed, following  \cite{Bambynek:1977zz}, as a sum 
over the possible levels of the  captured electron: 
\begin{eqnarray}
\lambda_{EC} &=& {G_{\beta}^2 \over {4 \pi^2}} \; \sum_i n_i \, C_i\,  \beta_i^2 B_i\times\nonumber\\
&& (Q-E_i)[(Q-E_i)^2-m_{\nu}^2]^{1/2} \; ,
\label{ECrate}
\end{eqnarray}
where $G_{\beta} = G_F \cos \theta_C$, $n_i$ is the fraction 
of occupancy of the $i$-th atomic shell, $C_i $ is the nuclear shape factor, 
$\beta_i$ is the Coulomb amplitude of the electron radial wave function 
(essentially, the modulus of the wave function at the origin) and $B_i$ 
is an atomic  correction for electron exchange and overlap. 
Note that  in eq.(\ref{ECrate}) there is a dependence on the neutrino mass for any single contribution in the sum.

\subsection*{Inner Bremsstrahlung end-point}

In beta decays, the neutrino mass can be measured because of the presence, 
in the rate, of the phase space factor for the antineutrino: 
$E_{\nu} \cdot p_{\nu} \simeq (Q-E_e) \cdot  \sqrt{(Q-E_e)^2-m_{\nu}^2}$.
An analogous factor appears in the expressions for the rate of IBEC 
(where the electron energy is replaced by the emitted photon energy) 
and of the EC decay with Auger electron emission. 
So far, only one experiment measured the $^{163}$Ho IBEC spectrum, 
but the sensitivity at the end-point was impaired by background \cite{IBEC_exp}.

\subsection*{Calorimetric absorption spectrum end-point}

In a calorimetric experiment the same neutrino phase space factor appears, with the total de-excitation energy replacing the energy of the electron. 
Although at a first glance  the calorimetric spectrum for EC decays 
may appear as a series of lines 
at energies $E_i$, where $E_i$ is the ionization energy of the 
captured electron, these lines have a natural width of a 
few eV and therefore the actual spectrum is a continuum with 
marked peaks (see Fig. 1). If the $Q$--value of the decay happens to be close to one of the $E_i$,  
the rate near to the endpoint, where the neutrino mass effects are relevant, will be greatly enhanced.

\begin{figure}[ht]
 \centering
  \includegraphics[clip=true,width=0.45\textwidth ]{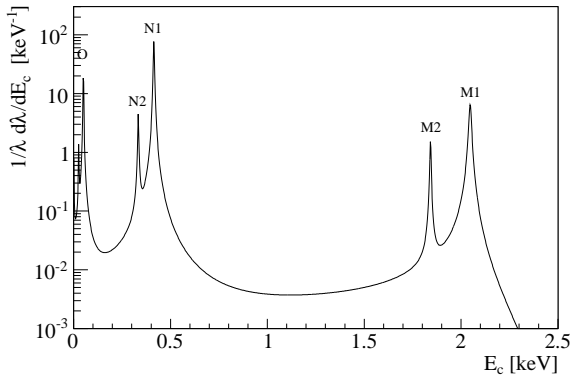}
 \caption{ Expected de--excitation energy spectrum of the  EC decay of $^{163}$Ho with $Q=2.5$~keV.  Detector resolution effects are not included. The parameters used in the calculation are given in Section \ref{section3}.
 }
 \label{fig:fig1}
\end{figure}

The distribution in de--excitation (calorimetric) energy $E_c$ is expected to be
\begin {eqnarray}
\label{E_c-distr}
{d \lambda_{EC}\over dE_c} &=& {G_{\beta}^2 \over {4 \pi^2}}(Q-E_c) \sqrt{(Q-E_c)^2-m_{\nu}^2} \;
\times  \\
&& \sum_i n_i  C_i \beta_i^2 B_i {\Gamma_i \over 2\,\pi}{1 \over (E_c-E_i)^2+\Gamma_i^2/4} \,, 
\nonumber
\end{eqnarray}
an expression derived in ref.\cite{DeRujula:1982qt}, where numerical checks to test the 
validity of the approximations made were also presented. 
As one can see, the very heights of the peaks in the energy distribution 
depend on the neutrino mass. The dependence is however too weak - for $Q$--values 
around 2.5~keV - to allow its determination. 

To date there are only three calorimetric absorption measurements 
reported in the literature:
\begin{enumerate}
\item{the ISOLDE collaboration used a Si(Li) detector with an implanted source \cite{Ravn:1984xj};}
\item{Hartman and Naumann used a high temperature proportional counter with organometallic gas \cite{HN};}
\item{Gatti et al. used a cryogenic calorimeter with a sandwiched source \cite{GattiHo:1997}.}
\end{enumerate}
However, none of these experiments had the sensitivity required for an 
end-point measurement and therefore they all gave results in terms of 
capture rate ratios. 
The most evident limitations of these experiments were statistics and 
energy resolution.
One further serious trouble for the Si(Li) and cryogenic detectors was 
incomplete energy detection caused by implant damages and  weak 
thermal coupling of the source, respectively.

Despite the shortcomings of previous calorimetric experiments, and
theoretical and experimental uncertainties, a calorimetric absorption 
experiment seems the only way to achieve sub-eV sensitivity for the 
neutrino mass.
Moreover, cryogenic microcalorimeters have reached the necessary maturity to 
be used in a large scale experiment with good energy and 
time resolution and are therefore the detector of choice for a 
sub-eV Holmium experiment.
 
\section{A roadmap for a Holmium Experiment}
\label{section2}

The road to achieve a sensitivity on the electron neutrino mass of 
0.1 eV/$c^2$ and beyond can be divided into two parts. The first one
is the demonstration of the feasibility of such an experiment, the second
one is the actual design and construction. In this section we discuss the 
steps necessary to demostrate the feasibility of a Holmium experiment,
while in the next we will focus on the actual experiment requirements.
We have identified four major steps necessary to demostrate the feasibility of 
a Holmium experiment, we discuss them in details in the following subsections.

\subsection{The $Q$-value of the decay}

The statistical 
sensitivity of the  $^{163}$Ho spectrum to the mass of the electron neutrino
is strongly dependent on the $Q$--value of the reaction. 
The smaller the $Q$--value (and thus closer to the energy of the M-peak)
the more sensitive the experiment will be.
As discussed before, at this time the $Q$--values obtained by means of the 
capture ratios are affected by large uncertainties both because of the 
error on the theoretical atomic physics factors involved and because of the 
correlation with the neutrino mass.
The various $Q$ determinations span from 2.2 to 2.8\,keV  and the 
current recommended value is 2.555$\pm$0.016 keV \cite{NDS}.

The first necessary step in the desing and implementation of a Holmium experiment
is an accurate, calorimetric measurement of the $Q$--value. 
The easiest, and most straightforward way to measure it is using a cryogenic
microcalorimeter similar to the one suggested for the full implementation of
the experiment.
Due to the looser constraints of the measurement, however, a single detector would 
be sufficient and the requirements on energy resolution could be significantly relaxed.
The statistics necessary to establih with high accuracy the end-point of the full calorimetric spectrum depends
on the $Q$-value itself. Using a detector with energy resolution of 10\,eV, about $5\times10^5$ counts would be
sufficient to determine the spectrum end-point with an accuracy better than 20\,eV for any $Q$-value between 2.2 and 2.8\,keV.
Notice that, for reasonable count rates (less than 100 counts/s) the measured $Q$-value
is not significantly affected by the detector pileup, so the total statistics
can be translated directly to a total time.

\subsection{$^{163}$Ho Production}

$^{163}$Ho was discovered at Princeton in 1968 in a sample of $^{162}$Er that was 
neutron activated in a nuclear reactor. Since its discovery, $^{163}$Ho has been 
produced in laboratory only for the purpose of nuclear and atomic property study. 
Since $^{163}$Ho  is not available off the shelf, a dedicated process must be set 
up to produce the amount needed by a neutrino mass experiment. 
The most challenging issue  is the achievement of the very high level of radio-purity 
required.
There are a few processes that are appealing for the intrinsic $^{163}$Ho 
production rate as compared to the accompanying production of radioactive 
contaminants. The most interesting ones are: 
\begin{itemize}
 \item neutron activation in nuclear reactor of 
$^{162}$Er [$^{162}$Er(n,$\gamma$)$^{163}$Er(75min)$\rightarrow ^{163}$Ho] 
with a cross section $\sigma$ of about 180\,barns;
 \item alpha particle bombardment of  $^{165}$Ho target 
[$^{165}$Ho($^4$He,*)$^{163}$Ho] with a $\sigma$ of about 0.05\,barns at 55\,MeV; 
 \item gamma bombardment of $^{165}$Ho target [$^{165}$Ho($\gamma$,2n)$^{163}$Ho] 
with a $\sigma$ of about 0.14\,barns at 17\,MeV.
\end{itemize}
The highest production rate is  achievable with the neutron activation of 
$^{162}$Er, even if to the detriment of contemporaneous built-up of very 
short half-life contaminants in the sample. The reactions with high energy 
projectiles have cross section thousands of times lower and therefore require 
higher intensity beams and longer activation times. On the other hand, they 
can use natural Holmium as target and have a highly suppressed contaminant production. 
Production of $^{163}$Ho  with the neutron activation of $^{162}$Er requires 
a detailed investigation of the process, because not all cross sections are 
fully known: a trade-off among reproducibility, easiness of the operations, 
production rate and radioactive background must be seeked.

A first very preliminary test of neutron activation of $^{162}$Er 38\% enriched 
sample of erbium oxide has been already done by some of the authors.
The irradiated sample, after waiting one month for the decay of the activity of 
short-lived isotopes, was first dissolved in a HCl solution and then deposited onto a 
Transition Edge Sensor (TES) microcalorimeter to be operated at 0.1\,K at the University of
Genova. The measured spectrum shows that $^{163}$Ho was indeed produced although 
it is impossible to estimate the background. The solution was analyzed by
Inductively Coupled Plasma Mass Spectrometry (ICP-MS) at the University of Milano-Bicocca and 
the presence of $^{163}$Ho was confirmed.

A large scale irradiation has also just been completed at the 1 MW 
Portuguese Research Reactor (RPI) at the Nuclear and Technological Institute, 
Instituto Superior T\'{e}cnico (IST-ID) in Lisbon. 
This reactor has a thermal column with average thermal and epithermal neutron 
fluxes of about $3\times10^{13}$~cm$^{-2}$~s$^{-1}$ and $3.7\times10^{11}$~cm$^{-2}$~s$^{-1}$, 
respectively. The estimated $^{163}$Ho production rate is about 
3~kBq/mg($^{162}$Er)/week (assuming the reactor is running 12 hours/day and 5 days/week). 

Purification and oxide reduction of Ho to metal can proceed as follows. 
First, three steps of cation-exchange column separation are carried out in order 
to absorb the rare earths and wash out the other impurities. Then, Ho is removed 
with Ethylenediaminetetraacetic acid (EDTA) and HNO$_3$ in multiple step processes. 
The resulting sample is mixed in a compounds for thermal reduction of Ho oxide. 
The final sample will be 
compound with metal Ho in low concentration, suitable as a target material for the 
following steps aiming at embedding the $^{163}$Ho in the detectors. 

\subsection{Detector Performance}

After 30 years from their introduction \cite{Moseley1984,Fiorini1984}, 
cryogenic microcalorimeters are a mature and reliable technology used in
many applications, including neutrino experiments.
Detectors with the proper energy resolution and speed for the proposed investigation
already exist, however, they must be designed and optimized for the requirements
of a Holmium experiment. We discuss here the principal characteristics of microcalorimeters
and their relevance to a Holmium experiment.
 
Microcalorimeters are composed of three parts, an absorber that converts 
energy into heat, a thermometer (or sensor) that detects the temperature variations of the 
absorber and a weak thermal link between the detector and a heat sink. 
When energy is released in the absorber, the temperature of the detector 
first rises and then returns to its original value due to the weak thermal 
link to the heat sink. The temperature change is proportional to the energy 
and is detected by the thermometer. 
Resistive thermometers (i.e., thermometers whose resistance changes with temperature)
have been the primary choice in many applications for years, but magnetic and kinetic
inductance thermometers have also, more recently, demostrated high 
performance \cite{TES,Silicons,KIDS,MAGCAL}.

In designing a neutrino experiment, three characteristics of microcalorimeters must be
considered: energy resolution, rise time of the events, and decay time. 
The effect of the energy resolution
on the outcome of a neutrino experiment is obvious (better resolution implies better 
accuracy), and it is quantified in section \ref{section3}. The energy resolution of 
a microcalorimeter is noise limited and independent of the energy being detected.
It depends on the heat capacitance $C$ of the detector, the temperature $T$,
and the sensitivity of the thermometer $\alpha$, according the the expression \cite{galeazzi2003}:
\begin{equation}
\Delta E \propto \sqrt{\frac{k_B CT}{\alpha}},
\end{equation}
where $k_B$ is the Botzmann constant. It is clear from the expression that, to obtain 
good energy resolution, it is important to work with small objects and material with
low specific heat (to keep the heat capacitance low), at low temperature, and with 
sensitive thermometers. Current technology has already achieved energy resolution below
2~eV~FWHM, using both TES and magnetic thermometers \cite{resolution}. 
Both technologies are currently considered for a Holmium experiment. However, to use them for the 
measurement of the electron neutrino mass, the $^{163}$Ho radioactive isotope must 
be embedded in the detector for a fully calorimetric measurement. Different procedures 
to embed the $^{163}$Ho in the detector are being investigated, and tests to 
verify that they do not significantly affect the detector performance are underway.

The rise time is the time it takes for the signal to change in response to the energy release.
This time is, in principle, the time needed for the thermometer temperature to rise and is
determined by heat diffusion in the absorber and the thermal coupling between absorber and 
thermometer. However, in many practical applications such time is quite fast and the rise
time is limited, instead, by the response time of the readout electronics. The event's rise time
is perhaps the most important parameter in a Holmium experiment as it affects directly the 
fraction of unresolved pileup events. If a second energy release happens within the rise of a previous
one, the two events may look like a single event of higher energy. The ensemble of such
events generates a second energy spectrum (unresolved pileup spectrum) which has
significant statistics in the region of interest for the neutrino mass ($Q$-value region), and
acts as a strong background component. Although the shape of the pileup spectrum can
be accurately modeled, the statistical uncertainty on the number of unresolved pileup events
can easily be the stronger limit on the neutrino mass sensitivity.
A significant effort is ongoing in order to develop analysis algorithms to identify and eliminate such
events. However, although we can now identify double events significantly closer than
the rise time \cite{armstrong2012}, in the end any Holmium experiment must be designed
with the smallest possible rise time to minimize pileup events. Rise times of a few 
micro-seconds are typical for TES microcalorimeters (limited by the readout electronics), 
while they can be significantly shorter for magnetic microcalorimeters.

The decay time is the time it takes for the signal to return to the initial level in order that
the detector be ready to register another event. An event happening on the decay of a previous one 
is usually easily identifiable, and does not contribute to the unresolved pileup spectrum.
However, due to detector's non-linearities, both events must be discarded, contributing to
the detector dead time. TES microcalorimeters have typical decay times below a millisecond,
making the dead time negligible for count rates up to hundred of counts per second. However, 
magnetic microcalorimeters could have much longer decay times, and the count rate may become an issue.
Magnetic penetration thermometers (MPT)\cite{shirron1993} use the same geometry as magnetic 
thermomenters but replaces the magnetic material with a superconductor in its transition. 
MPTs potentially combine the best of the magnetic calorimeter and TES technologies, providing 
the sensitivity of a TES in a dissipationless configuration \cite{bandler2012}.

Particular attention in a neutrino experiment must also be given to the readout electronics.
Both TES and magnetic microcalorimeters are read out using Superconducting Quantum
Interference Devices (SQUID). The direct approach of using individual 
readout channels for each detector has the necessary performance required by a Holmium
experiment. However, considering the number of detectors involved, such approach may not 
be practical. Fortunately, the same issues
affect several other investigations, and a massive effort on different SQUID multiplexing 
schemes is going on in laboratories around the World. These include time division multiplexing,
frequency division multiplexing, code division multiplexing, and microwave multiplexing
\cite{multiplexing}. Kinetic inductance detectors also offer a natual path to microwave
multiplexing.

\subsection{Theoretical Uncertainties} 

In spite of the many experiments, atomic and nuclear details of the
$^{163}$Ho decay are still affected by some uncertainty.
The natural width of the atomic levels involved in the 
atomic cascade following the EC is not directly known.
Moreover, the expected total absorption spectrum shape is not completely 
established. Riisager \cite{Riisager:1988wy} pointed out that the spectrum predicted in \cite{DeRujula:1982qt}  is only a first approximation because the 
rate at the end-point may be altered by the broadness of atomic level 
which affects the phase space available to atomic transitions.
It is, however, important to note that the $\it shape$ of the energy distribution 
near its endpoint is bound to be determined by the value of the 
neutrino mass, even if the rate itself may be uncertain because of the 
poor knowledge of atomic parameters. 

It  has also been observed \cite{Bennett3} that the ionization energies of interest in our 
equations differ from those obtained by Dysprosium excitation because of the presence of 
a further N$_{6,7}$ electron: those parameters will be measured with 
precision from the positions of the peaks in the $E_c$ distribution.
A high statistics and high energy resolution measurement of 
the total absorption spectrum is 
therefore a prerequisite for a neutrino mass measurement using the end-point as proposed in \cite{DeRujula:1982qt}.

\section{A Holmium Experiment}
\label{section3}

In this section we discuss how the statistical sensitivity depends
on the main experimental parameters. 
We then translate this in the experimental configurations required for a sub-eV sensitivity. 

The analysis is carried out using a frequentist Monte Carlo code
developed to estimate the statistical sensitivity of a neutrino mass
experiment performed with thermal calorimeters \cite{Nucciotti-Sens}. The approach is
to simulate the energy spectra that would be measured by a large
number of experiments carried out in a given configuration: the
spectra are then fitted as the real ones and the statistical sensitivity
is determined from the distribution of the obtained neutrino mass square,  $m^2_\nu$ \cite{Nucciotti-Sens}.

The Monte Carlo parameters for the simulated experimental configuration are the total statistics $N_{ev}$,
the FWHM of the detector energy resolution \de \fwhm\ and the fraction of unresolved pile-up events $f_{pp}$.
The simulated energy spectrum is then given by
\begin{eqnarray}
\label{eq:modelMC}
S(E) &=& {N_{ev}\over  \lambda_{EC}} \int_0^{2(Q-m_{\nu})} dE_1\, R(E-E_1)\,\cdot \nonumber \\
&& \left[{d \lambda_{EC}\over dE_c}(E_1) + {f_{pp}\over  \lambda_{EC}} \int_0^{E_1} dE_2 {d \lambda_{EC}\over dE_c}(E_2)\,\cdot \nonumber \right.\\ 
&& \left. {d \lambda_{EC}\over dE_c} (E_1-E_2)\right] \;,
\end{eqnarray}
where ${d \lambda_{EC} / dE_c}$ is given in (\ref{E_c-distr}) and the response function $R(E)$  is assumed to have a gaussian form
\begin{equation}
\label{eq:gauss} 
 R(E)=\frac{1}{\sigma \sqrt{2\pi}} \; \exp \big(-\frac{E^2}{2\sigma^2}\big)
\end{equation}
with standard deviation $\sigma = \Delta E_{\mathrm {FWHM}}/2.35$.
The input parameters are linked to the ones actually characterizing a real experiment: 
$N_{ev} = N_{det}  A_{\beta}  t_M$ and \mbox{$f_{pp} \approx A_\beta \tau_R$}, where
$N_{det}$ is the number of detectors, $A_{\beta}$ is the EC activity of a single detector, 
$t_M$ is the live time of the experiment, and $\tau_R$ is the pile-up resolving time (i.e., 
the minimum time separation between events 
that can be distinguished).

The free parameters in the fitting of the simulated spectra with Eq.~(\ref{eq:modelMC}) are the total statistics $N_{ev}$, the fraction of pile-up events  $f_{pp}$,
the EC transition $Q$--value and the squared neutrino mass  $m^2_\nu$.
The $^{163}$Dy atomic parameters $E_i$, $\Gamma_i$ and $\beta_i^2$ used for the numerical evaluation of (\ref{E_c-distr}) are considered known and therefore fixed to the values shown in Tab.\,\ref{En-Wid}. 
\begin{table}[h]
\centering
\caption{Energy levels of the captured electrons, with their widths, for $^{163}$Dy \cite{param1}.
Electrons squared wave functions at the origin $\beta_i^2$ relative to $\beta_{\rm M_1}^2$ \cite{Band:1985gm}.}
\begin{tabular}{cccc}
\hline
Level&$E_i$ [eV]&$\Gamma_i$ [eV] & $\beta_i^2/\beta_{\rm M_1}^2$\\
%&(eV)&(eV) \\
\hline
M$_1$ & 2047 & 13.2 &  1.0\\
M$_2$ & 1842 & 6.0 &  0.0526\\
N$_1$ & 414.2 & 5.4 & 0.2329 \\
N$_2$ & 333.5 & 5.3 & 0.0119 \\
O$_1$ & 49.9 & 3.0 &  0.0345 \\
O$_2$ & 26.3 & 3.0 &  0.0015 \\
\hline
\end{tabular}
\label{En-Wid}
\end{table}
The levels of the electrons that can be captured are fully occupied (i.e. $n_i=1$)
and the exchange and overlap 
corrections $B_i$
are neglected (i.e. $B_i\sim 1$).
\footnote{They are not given for all the levels needed in \protect\cite{Bambynek:1977zz}. Those given differ from unity by less than $\sim 10\%$.  The validity of this approximation far from the peaks  may be doubted, however the {\it shape} of the spectrum near to the end--point is anyhow determined by the neutrino phase-space factor. }

Given the large uncertainties on the  $Q$--value, the Monte Carlo study has been carried out for a set of reasonable values, i.e. 2200, 2400, 2600 and 2800 eV.

\begin{figure}[h!]
\begin{center}
% \vfill
 \includegraphics*[width=0.95\linewidth]{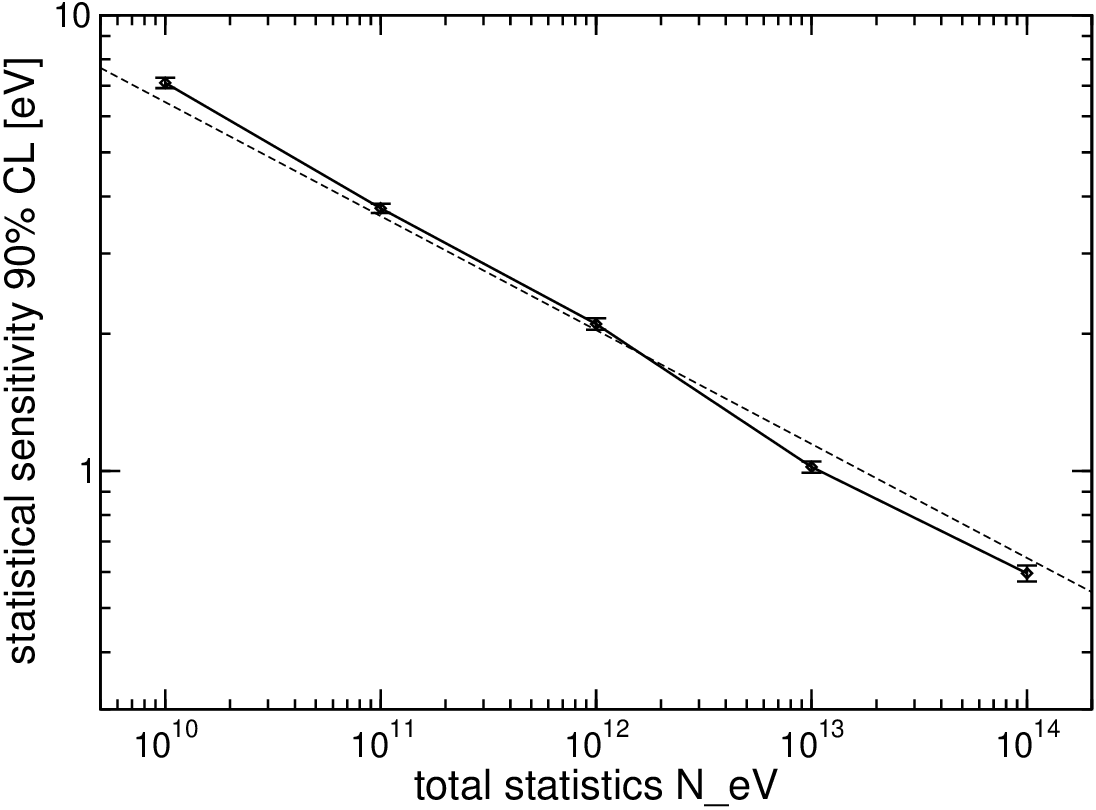}
\caption{\label{fig:stat} \de\fwhm$=1$\,eV, $f_{pp}=10^{-5}$, $Q=2600$\,eV.}
\end{center}
\end{figure}

From Fig.\,\ref{fig:stat} it can be appreciated how the total statistics $N_{ev}$ is crucial to reach a sub-eV neutrino mass statistical sensitivity. 
The fine dashed line on the plot corresponds to a $N_{ev}^{-1/4}$ functional dependence of the sensitivity as it would be naively expected for a $m^2_\nu$ sensitivity purely determined by statistical fluctuations. A more detailed analyis of the sensitivity as a function
of the statistics, for different pile-up fractions $f_{pp}$ and $Q$--values, shows that it can be interpolated as  $N_{ev}^{-1/\alpha}$, with $\alpha$ between 3.3 and 4.0: in the following we conservatively use $N_{ev}^{-1/4}$ to scale the Monte Carlo results.

\begin{figure}[h!]
\begin{center}
% \vfill
 \includegraphics*[width=0.95\linewidth]{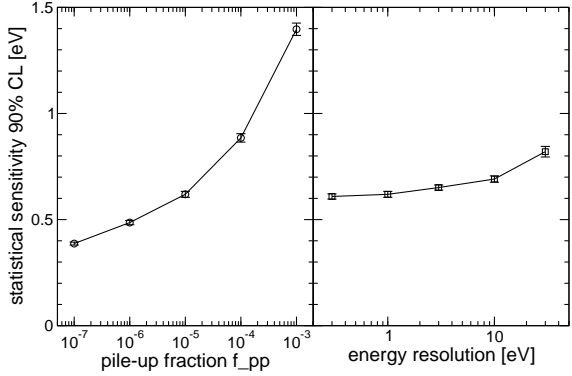}
\caption{\label{fig:de_fpp} $N_{ev}=10^{14}$, \de\fwhm$=1$\,eV, $Q=2600$\,eV (left). 
$N_{ev}=10^{14}$, $f_{pp}=10^{-5}$, $Q=2600$\,eV (right)}
\end{center}
\end{figure}
Fig.\,\ref{fig:de_fpp} shows the effect of both the energy resolution \de \fwhm\ and the 
fraction of pile-up events  $f_{pp}$ on the neutrino mass statistical sensitivity. 
As far as the energy resolution \de \fwhm\ is concerned, in the typical range accessible 
with today's microcalorimeter technology, its impact on the sensitivity of a Holmium experiment
is clearly small.

\begin{figure}[h!]
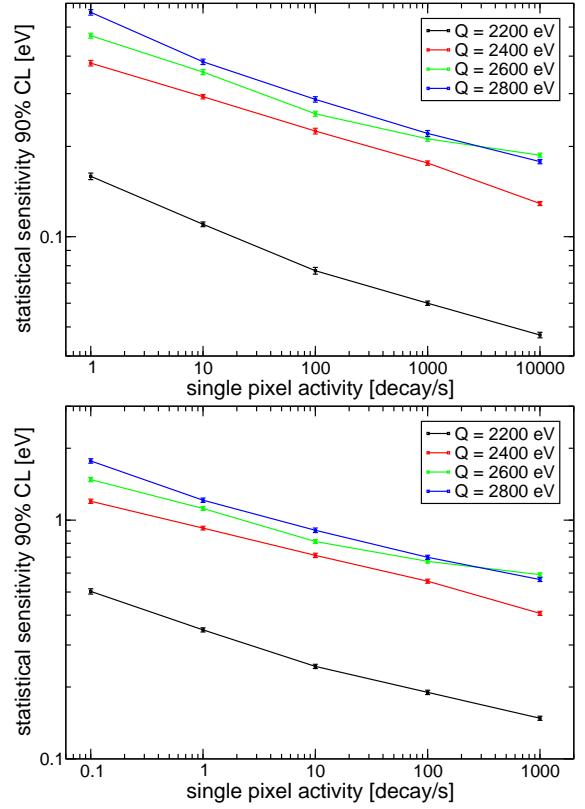

\begin{center}
 \includegraphics*[width=0.95\linewidth]{m.vs.A_de03tau01exp1e6.eps}
 \includegraphics*[width=0.95\linewidth]{m.vs.A_de1tau1exp1e5.eps}
\caption{\label{fig:activity} \de\fwhm$=0.3$\,eV, $\tau _R=0.1$\,\mus, $T=10^6$ detector$\times$year (top). 
\de\fwhm$=1$\,eV, $\tau _R=1$\,\mus, $T=10^5$ detector$\times$year (bottom).}
\end{center}
\end{figure}
On the contrary, the increase of  $f_{pp}$ has a strong effect on the
sensitivity. However it may pay out to increase the single pixel activity $A_{\beta}$, 
even if this entails 
an increased fraction of unresolved pile-up events  $f_{pp}$. 
This is demonstrated by Fig.\,\ref{fig:activity}, where the statistical sensitivity is plotted 
as a function of the single pixel activity $A_{\beta}$, for given pile-up resolving time $\tau_R$ and 
experimental exposure  $T=N_{det}t_M$ (with $N_{ev}=A_\beta T$).
Despite the higher fraction of unresolved pile-up events, the increase in pixel activity 
$A_{\beta}$ directly increases the experiment statistics for a given 
measuring time and the overall result is an improvement in the sensitivity of the experiment. 
It is worth recalling here that about $2\times10^{11}$ \Ho\ nuclei are needed for an activity 
of about 1 decay/s. 

\begin{figure}[hbt]
\begin{center}
% \vfill
 \includegraphics*[width=0.95\linewidth]{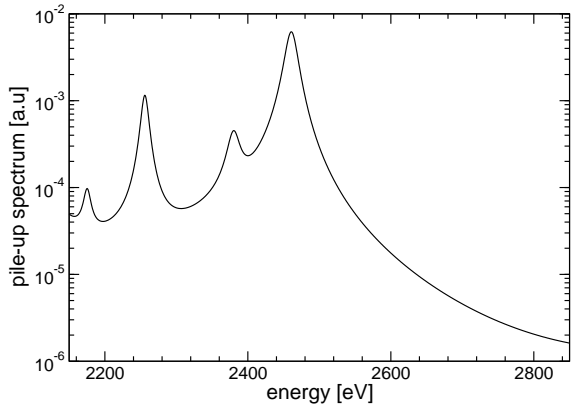}
\caption{\label{fig:pp} Pile-up spectrum.}
\end{center}
\end{figure}
Fig.\,\ref{fig:activity} also shows how the sensitivity depends on the transition $Q$--value. 
While the sensitivity is clearly much better for a $Q$--value of 2200\,eV, 
higher values result in quite similar sensitivities. In particular, because of the peculiar 
peaky shape of the pile-up spectrum in the 2200--2600\,eV interval (see Fig.\,\ref{fig:pp}), for high pile-up rates the sensitivity is
not monotonically improving when lowering the $Q$--value.

In Fig.\,\ref{fig:activity} two possible experimental configurations are considered. The lower panel refers to an experiment running arrays with today state-of-the-art thermal microcalorimeter technology: the experimental exposure is $10^5$\,detector$\times$year\, which, for example, could be achieved measuring for 10\,years arrays with a total of $10^4$ pixels.
The upper panel considers a more aggressive configuration with a factor 10 higher exposure and detectors with better energy and time resolution.

\begin{table}[]
\caption{\label{tab:sens} Experimental exposure required for various target statistical sensitivities, with $b=0$ and two
different sets of detector parameters. Configuration A is with \de\fwhm$=1$\,eV, $\tau_R=1$\,\mus\ and $A_\beta=1000$\,Hz.
Configuration B is with \de\fwhm$=0.3$\,eV, $\tau_R=0.1$\,\mus\ and  $A_\beta=10000$\,Hz.}
\begin{center}
\begin{tabular}{ccccc}
\hline 
$Q$ & target sensitivity &	\multicolumn{2}{c}{ exposure $T$ [detector$\times$year]}\\[0pt] % 
[eV] & [eV]   &	 Conf. A & Conf. B \\
\hline 
\hline 
2200 & 0.2 & 		$2.6\times10^{4}$& 		$3.3\times10^{3}$\\
2200 & 0.1 & 		$4.1\times10^{5}$& 		$4.8\times10^{4}$\\
2200 & 0.05 &		$6.6\times10^{6}$ &		$7.7\times10^{5}$\\
\hline 
2800 & 0.2 & 		$6.5\times10^{6}$& 		$6.3\times10^{5}$\\
2800 & 0.1 & 		$1.0\times10^{8}$& 		$1.0\times10^{7}$\\
2800 & 0.05 &		$1.7\times10^{9}$ &		$1.6\times10^{8}$\\
\hline 
\end{tabular} 
\end{center}
 \end{table}
To conclude, in Tab.\,\ref{tab:sens} we have evaluated the experimental exposure required to reach target neutrino mass statistical sensitivities of 0.2, 0.1 and 0.05\,eV for two different sets of detector parameters (comparable to those in Fig.\,\ref{fig:activity}) and for
the two extremes of the $Q$--value range.

\begin{figure}[hbt]
\begin{center}
% \vfill
 \includegraphics*[width=0.95\linewidth]{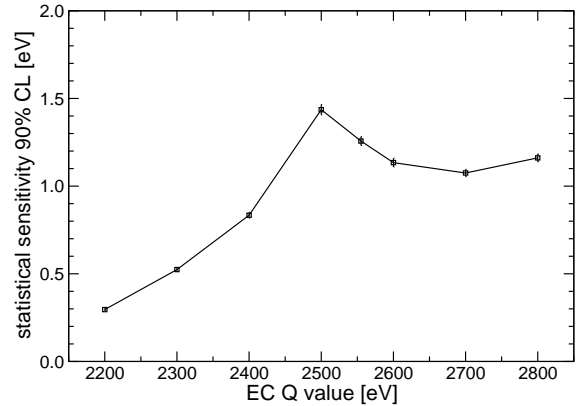}
\caption{\label{fig:olmio1} Pilot experiment sensitivity.}
\end{center}
\end{figure}
On a short time scale, a smaller size pilot experiment may be carried out with the aim of testing 
the potential of a Holmium experiment. Fig.\,\ref{fig:olmio1} shows the \mn\ statistical sensitivity
achievable for total statistics of about $8.5\times10^{13}$\,events, using detectors with an energy and time resolution
of 1.5\,eV and 1\,\mus, respectively, each with an \Ho\ activity of about 300 decay per second. This exposure could be obtained for example by running 3000 detectors for about 3 years. The sensivities in Fig.\,\ref{fig:olmio1} have been estimated for  $Q$--values in the 2200--2800\,eV range and span from about 0.3 to about 1.5\,eV.

\section{Conclusions}
The absolute value of the neutrino masses is not yet known. The most stringent upper limits on them  
come from experiments on tritium beta decay with electrostatic spectrometers: the best limit of 
sensitivity to which this technique can be pushed in the future  seems to be $\sim$0.2 ${\rm eV}$, 
as in the KATRIN experiment \cite{KATRIN}. A completely different technique  using cryogenic 
microcalorimeters to perform a calorimetric experiment has been proposed and applied in pilot 
experiments measuring the beta decay of the isotope $^{187}$Re, which has a very low Q--value 
but also a very long lifetime, and a large experiment, MARE \cite{MARE}, will follow this path. 

The technique has  in fact reached a maturity that allows to envisage a full scale experiment in 
order to reach an even better sensitivity. In this paper, we have presented the roadmap to 
reach this goal in an calorimetric experiment using the electron capture in $^{163}$Ho - 
an isotope having similar Q--value but much shorter lifetime - to obtain a limit on the 
electron neutrino mass, following a suggestion made many years ago \cite{DeRujula:1982qt}. 
This paper shows that such an experiment is clearly challenging, but doable, and 
we identified a clear path and technical steps necessary for its success.

\section*{Acknowledgements}
The authors would like to thank Dr. Ezio Previtali, Dr. Massimiliano
Clemenza, and Dr. Andreas Kling for their support. 

\bibliographystyle{elsarticle-num}

\end{document}